# PET Synthesis via Self-supervised Adaptive Residual Estimation Generative Adversarial Network

Yuxin Xue, Lei Bi, Yige Peng, Michael Fulham, David Dagan Feng, *Fellow, IEEE*, Jinman Kim, *Member, IEEE*

*Abstract*— Positron emission tomography (PET) is a widely used, highly sensitive molecular imaging in clinical diagnosis. There is interest in reducing the radiation exposure from PET but also maintaining adequate image quality. Recent methods using convolutional neural networks (CNNs) to generate synthesized high-quality PET images from 'low-dose' counterparts have been reported to be 'state-of-the-art' for low-to-high image recovery methods. However, these methods are prone to exhibiting discrepancies in texture and structure between synthesized and real images. Furthermore, the distribution shift between low-dose PET and standard PET has not been fully investigated. To address these issues, we developed a self-supervised adaptive residual estimation generative adversarial network (SS-AEGAN). We introduce (1) An adaptive residual estimation mapping mechanism, AE-Net, designed to dynamically rectify the preliminary synthesized PET images by taking the residual map between the low-dose PET and synthesized output as the input, and (2) A self-supervised pre-training strategy to enhance the feature representation of the coarse generator. Our experiments with a public benchmark dataset of total-body PET images show that SS-AEGAN consistently outperformed the state-of-the-art synthesis methods with various dose reduction factors.

## I. INTRODUCTION

POSITRON Emission Tomography (PET), an ultrasensitive and non-invasive molecular imaging technique, is considered as the main imaging instrument for oncology [1], neurology [2], and cardiology [3]. Compared with other imaging modalities such as Magnetic Resonance Imaging (MRI) and Computed Tomography (CT), PET can image the functional properties of living tissue and detect disease-related functional activity within organs by injecting radioactive tracers into the body [4]. Unfortunately, the ionizing radiation dose from the injected radioactive tracer increases the chance of patients' radiation exposure and therefore limits its application [5]. According to the dose level of the tracer, the reconstructed PET images can be classified as standard-dose PET (*sPET*) that refers to the commonly used imaging protocol for PET scans, which typically involves injecting a patient with a radiotracer dose of about 3-5 mCi (millicuries) and acquiring images over a period of 30-60 minutes and low-dose PET (*lPET*) images that uses a lower radiation dose compared to standard-dose PET. These 'high quality' *sPET* images contain better structural details and have higher signal-to-noise (SNR) ratios of the radiotracer distribution when compared with the *lPET* counterpart. However, *sPET* requires higher cumulative radiation exposure, which thereby raises potential health risks, resulting in restricted usage e.g., among children [48]. Motivated by these challenges, there have been great research interests in developing new image analysis methods that allow to reconstruct *sPET* images using *lPET* images [9]-[13]. One category of studies has been focusing on reconstructing high-quality PET from sinogram data with low dose sinogram [9-11]. Although these algorithms achieved good results, their applicability may be constrained due to the slow convergence and longer scan time. An alternative is to reconstruct *sPET* as a post-reconstruction process from *lPET*. Denoising methods, including combining complementary wavelet and curvelet transforms [12], and non-local means [13], were utilized to improve *lPET* image quality. However, these methods aimed at reducing the noise of *lPET* by removing unwanted distortions such as random noise, artifacts, and interference.

Deep learning methods based on convolutional neural networks (CNNs) have achieved great success in medical image analysis-related tasks e.g., automated tumor segmentation and classification [14]-[16]. Motivated by the great capacity of feature representation from CNNs, investigators have attempted to use it for reconstructing *sPET* from *lPET* images. Xiang *et al*. [16] integrated multiple CNN modules following an auto-context strategy to estimate *sPET* from *lPET*. Kim *et al*. [17] proposed a local linear fitting (LLF) function and a denoising convolutional neural network (DnCNN) to enhance image quality from *lPET*. Spuhler *et al*. [18] adopted a U-Net architecture and replaced commonly used convolutional kernels with dilated kernels to increase the receptive field.

To address these challenges, investigators used generative adversarial network (GAN) to preserve detailed information. GAN extends the CNNs by adding an additional discriminator network to distinguish between real/synthetic images [19]. Bi *et al*. [20] developed a multi-channel GAN to synthesize high quality PET. Wang *et al*. [21], leveraged a conditional GANs

Manuscript submitted April 18, 2023. (Corresponding author: Jinman Kim.)
This work did not involve human subjects or animals in its research.
Y. Xue, Y. Peng, D. Feng, and J. Kim are with the School of Computer Science, University of Sydney, NSW, Australia. Y. Xue (E-mail: yxue8704@ uni.sydney.edu.au); Y. Peng (E-mail: ypen6619@ uni.sydney.edu.au); J. Kim (E-mail: jinman.kim@sydney.edu.au).
Lei Bi is with the Center for Translation Medicine, Shanghai Jiao Tong University, Shanghai, China (E-mail: lei.bi@sjtu.edu.au).
M. Fulham is with the Department of Molecular Imaging, Royal Prince Alfred Hospital, Camperdown, NSW, Australia, and also with the Sydney Medical School, University of Sydney, Camperdown, NSW, Australia (E-mail: michael.fulham@sydney.edu.au).
D. Feng is also with Med-X Research Institute, Shanghai Jiao Tong University, China (E-mail: dagan.feng@sydney.edu.au).

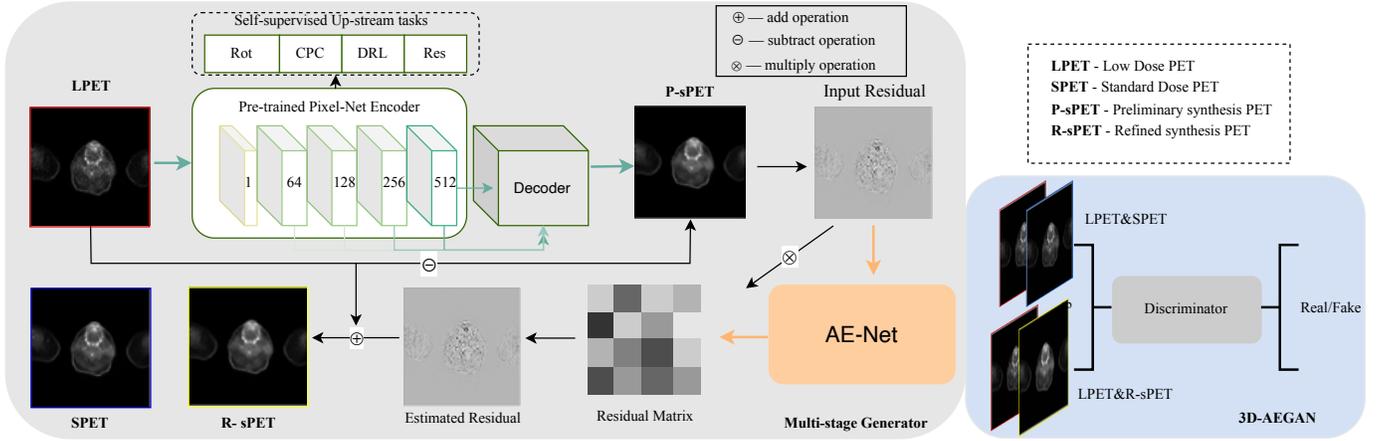

Fig. 1. Our proposed SS-AEGAN framework to synthesize high-quality *sPET* from *lPET*.

model and an adversarial training strategy to recover *sPET* images from *lPET*. A GAN-based model with feature matching and task-specific perceptual loss was proposed by Ouyang *et al*. [22] to accurately yield *sPET* from *lPET* counterpart. Unfortunately, these GAN-based approaches have difficulties in recovering high dimensional details, e.g., contextual information, and clinically significant texture features e.g., intensity values of an image. This is mainly attributed to the fact that they have not explored the spatial distribution correlations between the *sPET* and *lPET* images. Furthermore, existing GAN methods, when applied to PET images, tend to produce artifacts due to the use of a sequential up-sampling process. To compensate for the unsatisfied synthesized PET results from the GAN model, Luo *et al*. [33] adopted a residual estimation module to predict the residual between the preliminarily synthesized PET and the real PET images. However, the residual estimation network only accounted for the preliminary synthesized results as the input such that the quality of the residual mapping was heavily reliant on the preliminary synthesized PET. In addition, [33] proposed a 2D-based synthesis model that may not fully use the spatial dependencies and structural information along the three views.

Besides, the existing methods have not explored to employ self-supervised pre-training to improve the generalizability of synthesis model which can facilitate domain adaptation by learning representations that are invariant to differences in imaging protocols, scanner types, and other acquisition variables.

In this study, we propose a self-supervised adaptive generative adversarial network (SS-AEGAN) to recover *sPET* images from corresponding *lPET* images. When compared to the state-of-the-art methods, we introduce the following contributions:
1) A 3D input-involved adaptive residual estimation module, AE-Net, by taking the difference map between the preliminary *sPET* and the *lPET* as the input, such that it allows the network to learn the residual mapping between the *sPET* and the *lPET*. When compared with the existing residual learning methods, our *lPET*-embedded rectification can potentially correct the artifact brought by the initial results.
2) A self-supervised pre-training strategy to initialize the encoder of the coarse synthesis generator with multiple upstream tasks to aid the downstream task – high-quality *sPET* synthesis. Our proposed multiple upstream tasks allow for enhancement of the feature representation capability such that the context and structural information of the synthesized images can be iteratively reconstructed at a finer convolutional layer.

This work is an extension of our preliminary work where we introduced the ability to reconstruct high-quality PET from low-dose counterpart via a classification-guided generative adversarial network with super-resolution refinement (CG-SRGAN) [51]. We have made the following additional contributions to improve the overall performance and overcome its limitations: (a) when compared to CG-SRGAN, our new SS-AEGAN innovates by integrating an adaptive residual estimation mechanism - AE-Net and self-supervised pre-training heads; (b) Our proposed SS-AEGAN differs from CG-SRGAN in that it employs a new residual estimation learning approach to overcome the inherent dependency on input PET image quality. This approach can refine the synthesized output without relying heavily on the quality of the input PET images. Furthermore, we included a self-supervised pre-training step to enhance the generalizability of our model, thereby improving its effectiveness in synthesizing PET images using three different tracers across two scanners; and (c) we conducted additional experiments with total-body PET images acquired from the Siemens scanner to validate the cross-scanner generalizability of the proposed SS-AEGAN.

## II. RELATED WORK

Our work is closely related to three tasks in medical image analysis, which are (a) PET image synthesis, (b) residual estimation and (c) self-supervised pre-training.

### A. PET Image Synthesis

PET image synthesis involves the reconstruction of high-quality PET images from denoised PET data. U-Net-based structures are commonly used backbones. Xu *et al.* [22] proposed a modified 2.5D U-Net model which takes multi-slice PET as input to reconstruct *sPET*. Chen *et al.* modified 3D-UNet by adding pixel unshuffled and shuffle layer on the first encoder block and last decoder block to obtain high-quality PET which won first place in the 2022 MICCAI challenge [23]. Inspired by the success achieved by the GAN model [19], investigators have also attempted to adopt GAN for PET image synthesis. Wang *et al.* [21] proposed 3D-cGAN to generate *sPET* by taking 3D low-dose PET as input. Wang *et al.* [24] stacked multiple 3D-cGAN together to build stack-GAN for high-quality PET synthesis which proved that multi-GAN structures can outperform single 3D-cGAN. To explicitly encourage embedding the semantic information of PET images to latent space, Cycle-GAN has been widely applied to *sPET* synthesis [25][26] which combines three types of loss: adversarial loss, cycle-consistency loss, and identity loss to train two pairs of generator and discriminator.

In this study, we adopted a 3D-GAN-like structure as the backbone, which utilizes a modified 3D-UNet as the generator. This approach was chosen due to the strong capability of the modified 3D-UNet to capture both global and local features. Different from the existing methods, we employ a self-supervised pre-training strategy to enhance the feature generalization capability of the generator so that synthetic PET can retain more semantic and structural features.

*B. Residual Estimation Mapping*

Residual learning [27] is initially incorporated into CNNs to optimize the training process. It demonstrates that adding residual block into the CNNs can effectively address the degradation problem on various image processing tasks e.g., image super-resolution [28], image denoising [29][30], and so on. Nie et al. [31] applied long-term residual connection to the generator by adding a skip connection from the input to the final layer and then performing an element-wise addition on the 3T MRI to 7T MRI synthesis task. However, the above work leveraged residual learning for the purpose of assisting the training process and preventing gradients from vanishing. In addition, residual learning can also be used to identify the connections between the input image and the target image on the synthesis task. Wu *et al.* [32] developed a residual learning structure by employing convolution layers to fuse images, which improved the cross-modality synthesis task. Despite applying this kind of skip-connection-based residual learning that can bridge the gap between the input and the target, it still fails to learn the residual mapping between them. Multiplicative residual scheme has been leveraged for attenuation corrected PET by setting divided mapping between input (uncorrected PET) and target (corrected PET) as the learning objective [34]. AR-GAN [33] proposed by Luo *et al.* first proposed a separate residual estimation network that took the synthetic PET from the previous generator as the input to predict residual mapping between the *lPET* and the *sPET*.

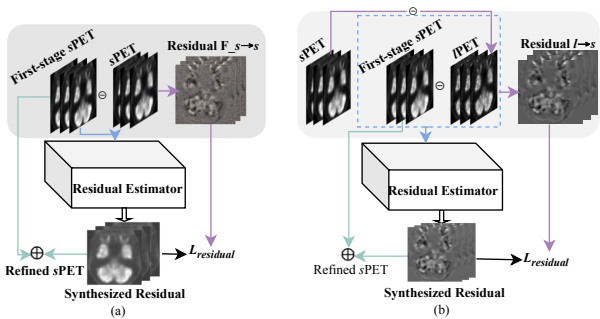

**Fig. 2.** Two residual learning mechanisms are distinguished from the input of residual estimator and target residual mapping. (a) proposed by [33] in which input is the first-stage synthetic *sPET* with target residual mapping of first stage *sPET* to real *sPET*. (b) proposed *lPET* involved residual learning method, where input is a residual between *lPET* and the first stage *sPET* with target residual mapping of *lPET* to real *sPET*.

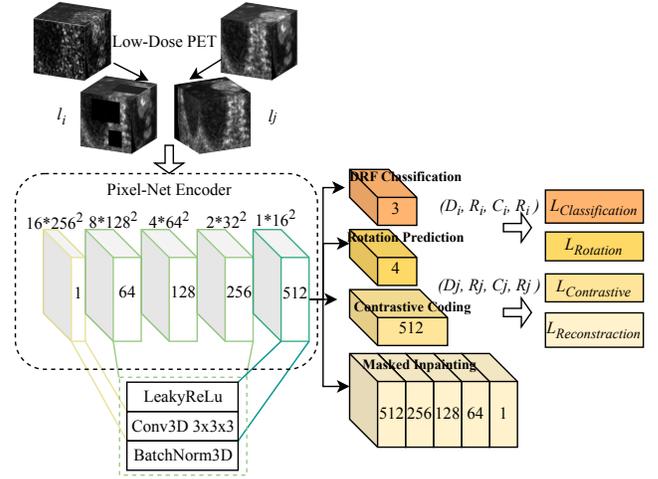

**Fig. 3.** The self-supervised pre-training strategy for 3D-AEGAN. The *lPET* with DRF4 to DRF100 are pre-processed by random rotation and mask operation. The augmented data are sent into the Pixel-Net encoder for the up-stream tasks: DRL classification, rotation angle prediction, contrastive coding, and self-restoration to learn more comprehensive content image features.

Our proposed method extends the definition of residual estimation mapping with: (1) when compared to the commonly used 2D-based approaches, we introduce a 3D-based residual estimation approach that ensures all the spatial and contextual information can be captured; and (2) we adopt the input guided residual estimation strategy to overcome the artifacts brought by the GAN structure that allows to further boost the PET synthesis outcomes while producing trustworthy results.

*C. Self-supervised Pre-training*

Self-supervised learning aims to achieve supervised feature learning where the tasks for the supervision are produced by the data itself. There are various self-supervised learning strategies for medical images. One is the prediction of relative positions (PR) between patches [35] which is motivated by the intrinsic position relations among the divided parts of an object of interest. Another example of a PR task is image rotation prediction [40]. Because the PR method is based on patches that do not learn global context representation, it can only provide limited improvements for the tasks that require global context such as classification. Image context recovery based on self-supervised learning shows better feature representation ability. The idea is to train CNNs to 'inpaint' missing information in the images with randomly removed patches [36][37]. Recently, contrastive coding has been adopted in self-supervised pre-training (SSP) which learns high-dimensional shared features by maximizing the mutual information from the encoded representation of positive image pairs [38][39]. Inspired by [41], three sub-tasks, including rotation prediction, contrastive coding, and image inpainting, drive our SSP method. The pre-trained head is embedded with the encoder of the generator to boost feature extraction capability for the subsequent PET synthesis task.

## III. METHODS

The framework of our proposed SS-AEGAN is shown in **Fig.1**. It consists of three components: a) a synthesis network – Pixel-Net to generate an initial prediction result that closely resembles the actual *sPET* image; b) a refinement network - AE-Net to estimate the residual between *sPET* and *lPET* by taking residual mapping of the difference between first stage result and *lPET*; and c) a discriminator to distinguish the veracity of the refined PET. SS-AEGAN is trained by two stages: one is self-supervised pre-training on the encoder of Pixel-Net by four upstream tasks; the second stage is to train the whole synthesis model in an end-to-end manner. Specifically, the adjacent AE-Net uses the residual map of the preliminary *sPET* and *lPET* as input to learn adaptive residual parameters after Pixel-Net first generates preliminary estimated *sPET* (P-*sPET*) from *lPET*. After that, the estimated residual can be calculated by multiplying the output of the AE-Net with the input residual map. As a result, the estimated residual and preliminary Pixel-Net output is incorporated into the final refined *sPET*. Finally, the rectified synthetic/real *sPET* image and the associated *lPET* image are sent into the discriminator, which is then trained to distinguish between the actual and synthetic image pairs.

### A. Coarse Generator Network – Pixel-Net

Due to the strong ability to capture the texture and semantic features, U-Net [43] has become a commonly used network as an image synthesis generator. Therefore, a 3D U-Net-based network Pixel-Net is applied as a coarse generator to reconstruct *sPET* from *lPET* at the pixel level. The encoder is composed of five down-sampling blocks, and each of them adopts a 3x3x3 convolutional kernel with stride 2. The encoder blocks are introduced in the form of LeakyReLu – Convolutional layer – Batch Normalization (BN). The decoder structure also contains five up-sampling blocks with convolutional kernel 3x3x3 and stride 2. The decoder blocks are made up of three sequential parts: ReLu, a transposed convolutional layer, and BN. Note that, the first encoder block only contains one convolutional layer itself and the last block of both the encoder and decoder removes BN operation to preserve more original texture and structure details for better synthesis output. Besides, the skip connection of U-Net is also utilized in our Pixel-Net to efficiently replenish the low-dimensional information that could be lost during the up-sampling process between the encoder blocks and the related decoder blocks.

### B. Adaptive Residual Estimation – AE-Net

Due to the limited synthesis ability of Pixel-Net, the preliminary *sPET* is still different from the true *sPET* in local texture and global structure. To address this problem, we propose an adaptive residual estimation network – AE-Net to refine the first-stage synthesis results from Pixel-Net, inspired by the AR-Net of [33]. **Fig.2** illustrates two residual mapping strategies: AR-Net [33] and the proposed AE-Net which also compares the input of the estimator and target residual mapping. AR-Net assumes that more realistic *sPET* can be obtained by incorporating the residual to preliminary prediction so that the residual estimator takes preliminary prediction as input to generate a residual mapping between real *sPET* and preliminary prediction shown in **Fig.2a**. However, our hypothesis is that residual mapping of preliminary synthetic *sPET* and *lPET* can better incorporate shared data distribution between *sPET* and *lPET*. **Fig.2b** shows the residual estimation pipeline of AE-Net which takes the residual map between the prediction result of Pixel-Net and *lPET* as input to generate a refined residual matrix as output. The target residual map is obtained by multiplying input with the refined output matrix, then the final reconstructed *sPET* is the sum of the residual map and preliminary *sPET*. AE-Net is a symmetrical V-shaped network composed of eight encoder/decoder blocks with Conv - BN - Leaky-ReLU/ReLU components and all the convolutional layers apply 3x3x3 kernels. For the encoder section, the down-sampling is accomplished by Max-pooling operations with a stride of 2. The decoder part uses transposed convolutions with a 3x3x3 filter of stride 2 for up-sampling.

### C. Discriminator

Pixel-Net and AE-Net jointly operate as the generator in the GAN model. Another crucial component of GAN, the discriminator network, seeks to differentiate between the synthesized *sPET* and target *sPET*. The discriminator takes either the fake image pair of synthesized *sPET* and *lPET* or the real image pair of *sPET* and *lPET* as input and differentiates whether the input is real or fake. When the generator and discriminator engage in adversarial learning, the synthesis results are improved by Pixel-Net and AE-Net, and the discriminator is boosted in the opposite direction to identify the real/fake image pair. The discriminator contains five blocks in the form of Conv - Leak ReLU - BN with the last block's activation function replaced by sigmoid.

### D. Self-supervised pre-training strategy

To maximize the feature representation ability, we applied a self-supervised pre-training strategy on the encoder of Pixel-Net with four upstream tasks. Dose reduction level (DRL) classification, rotation prediction, and contrastive coding tasks with the objective of learning global features e.g., anatomical information. PET restoration task, on the other hand, aims to learn context-level features by inpainting perturbated images. As illustrated in **Fig.3**, mixed *lPET*s will be randomly augmented to produce sub-patches with two perspectives, then projected into latent embedding space by Pixel-Net encoder in parallel. For the up-scream tasks, the encoder is coupled to four sub-branch heads. Specifically, for the DRL classification task, the sub-branch head predicts the DRF label of input *lPET* by impacting multi-scale features to a sequence of linear layers. Rotation angle classification is also utilized for representation learning with the objective of predicting the rotated degree of augmented *lPET* in which high-dimensional feature representation extracted from Pixel-Net encoder was projected to a four-dimension vector by Identity-Linear operation head. To encode more underlying shared information of high-dimensional features and eliminate low-level noise, contrastive predicting coding (CPC) head is used to maximally preserve mutual information of perturbated 3D patches from the same *lPET*. The CPC branch has a similar structure to the Rotation branch, but with 512 channels output instead. The anatomical pattern for organ shape, edge information, and texture are also expected to be displayed through encoded features. By integrating a decoder with a Pixel-Net encoder, the self-restoration branch is employed to learn anatomically related representation with the goal of inpainting cut-off *lPET* patches. The structure of self-restoration branch is consistent with the structure of the proposed Pixel-Net.

*E. Objective Functions*

The data flow of our proposed SS-AEGAN is from Pixel-Net to AE-Net, then to discriminator, each of which would be optimized by a loss function. The self-supervised pre-training strategy is applied before SS-AEGAN is trained in an end-to-end manner.

*a) Self-Supervised Pre-training for Pixel-Net Encoder*

*lPET* from training datasets is divided into three classes according to their dose reduction level: $k_1$, *lPET* with DRF 4 and DRF 10; $k_2$, *lPET* with DRF 20 and DRF 50; $k_3$, *lPET* with DRF 100. The category prediction $\hat{y}$ is yielded by DRL classification branch. It is trained with penalty of cross-entropy between ground-truth $y$ and prediction:

$$L_{Classification} = -\sum_{k_i}^{K} y^{(k_i)} \log \hat{y}^{(k_i)} \quad (1)$$

During SSP, the input *lPET* will be augmented by random rotation operation with a rotating angle of 0°, 90°, 180°, and 270°. The rotation prediction branch will project encoded features f into prediction probability $\hat{p}$ for each class c by rotation head. Multi-class cross-entropy loss is used to regulate the training process as follow:

$$L_{Rotation} = -\sum_{c}^{C} y^c \log \hat{p}^c \quad (2)$$

Where $y$ indicates the real rotation angle of input instance.

Contrastive predicting coding (CPC) branch is expected to be fruitful for learning the high-level shared information between the augmented patches from the same source data. Encoder $G_{enc}$ maps input volume $x_i$ into a latent representation $z_i = G_{enc}(x_i)$.

Further projecting $z_i$ into a latent context space with CPC head $G_{cpc}$ results in constative coding representation $c_i = G_{CPC}(z_i)$. Mutual information between a pair of representations is measured by the cosine similarity distance (*CS*). CPC loss is used to maximize the mutual information of positive pairs $c_i$ and $c_j$ which are augmented from the same input and minimize the mutual information of negative pairs $c_i$ and $c_k$ which have different views. Overall, the CPC loss is defined as:

$$L_{CPC} = -\log \frac{\exp\left(\frac{CS(c_i, c_j)}{\sigma}\right)}{\sum_{k}^{2N} I_\Omega \exp\left(\frac{CS(c_i, c_k)}{\sigma}\right)} \quad (3)$$

where $\sigma$ is the normalization scale and $I_\Omega$ is the indicator function with sample space $\Omega = \{k \neq i\}$. N denotes training batch size.

The self-restoration branch is used to generate the masked patches. During augmentation, a cut-out operator $\psi$ is applied on input 3D *lPET* volume $v$ to obtain perturbed patches $\tilde{v} = \psi(v)$. Cut-out operator $\psi$ includes operations of random dropout 30% volume, local-shuffling, and out-painting which are proposed by [44], [45]. The self-restoration branch is optimized by minimizing the L1 loss between the input volume and restored output:

$$L_{Res} = \frac{1}{N} \sum_{i}^{N} \|v_i - \mathcal{F}(\tilde{v}_i)\|_1 \quad (4)$$

where N is batch size and $\mathcal{F}$ represents the self-restoration process.

The above four branch heads are integrated with a shared encoder to be optimized by the total self-supervised pre-training loss function:

$$L_{SSP} = \lambda_1 * L_{Classification} + \lambda_2 * L_{Rotation} \\ + \lambda_3 * L_{CPC} + \lambda_4 * L_{Res} \quad (5)$$

*b) Loss Function for 3D-AEGAN*

We modify the objective function beyond the standard adversarial loss of a GAN to incorporate voxel-wise content loss along with image-wise loss, to ensure spatial alignment of the enhanced full dose images with the ground truth.

Specifically, the input low-dose PET $V_L$ will be sent into the first-stage generator Pixel-Net $P$ to generate preliminary synthesis result $P(V_L)$. To enforce the Pixel-Net to predict results aligned with real *sPET* at the pixel level, image content loss is introduced, which is formulated as follows:

$$\mathcal{L}_{content} = \mathbb{E}_{V_L \sim P_L}[\|V_S - P(V_L)\|_1] \quad (6)$$

The target residual $r$ is obtained by elementwise subtracting the real *sPET* $V_S$ from input *lPET* $V_L$. The input of AE-Net $R$ is the difference map $\tilde{r}$ between preliminary result $P(V_L)$ and $V_L$, and the output is the adaptative residual parameters matrix $R(\tilde{r})$. The estimated residual map is then produced by performing an element-wise multiplication of the input residual with the residual parameter matrix. As the regularization term, an L1 loss is used to punish residual error, which is defined as follows:

$$\mathcal{L}_{residual} = E_{\tilde{r} \sim P_{\tilde{r}}}[\|r - R(\tilde{r}) * \tilde{r}\|_1] \quad (7)$$

The estimated residual and corresponding *lPET* are incorporated to generate the final synthesis results. To further improve image quality, we applied an additional adversarial objective function, which was defined as follows:

$$\mathcal{L}_{adv} = E_{V_L \sim P_L, V_S \sim P_S}[(D(V_L, V_S) - 1)^2] \\ + E_{V_L \sim P_L}[D(V_L, R(\tilde{r}) * \tilde{r} + V_L)^2] \quad (8)$$

The final objective function of our proposed 3D-AEGAN consists of three types of loss function: content loss ($\mathcal{L}_{content}$), residual loss ($\mathcal{L}_{residual}$) and adversarial loss ($\mathcal{L}_{adv}$). Thus, the overall loss function was described as:

$$\mathcal{L}_{total} = \lambda_5 \mathcal{L}_{content} + \lambda_6 \mathcal{L}_{residual} + \lambda_7 \mathcal{L}_{adv} \quad (9)$$

*F. Implementation Details*

Due to the differences between the two PET scanners in spatial resolution and data distribution, we trained and tested two datasets separately. Specifically, each dataset was randomly divided into training, validation, and test sets with a ratio of 0.8: 0.1: 0.1, respectively. We used overlapping patches from *lPET* and *sPET* to reduce computational costs. For supervised learning, *lPET* and *sPET* images were arbitrarily cropped into patches of 256 × 256 × 16. ALL the PET scans were converted into SUV to normalize the images. The final recovered images were obtained by merging the overlapping patches.

Self-supervised pre-training was applied on the encoder of the Pixel-Net with four sub-branches by using a batch size of 4 and AdamW optimization [46]. The hyperparameters $\lambda_{1-4}$ in Eq. (5) were set to 1 empirically.

The 3D-AEGAN was trained using a batch size of 4 with the Adam optimization [47]. We empirically set $\lambda_5 = 300$, $\lambda_6 = 10$, and $\lambda_7 = 1$ for the hyperparameters defined in Eq. (4) and were fixed in the subsequent tests. We trained the proposed method for 100 epochs The learning rate was initially set as 2e-4, which was then linearly decreased with a factor of 0.1 and patience of 5 epochs. To avoid overfitting, an early stopping strategy was applied and was used to terminate the training process when the learning rate exceeds 2e-6. All the experiments were conducted on an 11GB NVIDIA GeForce RTX 2080Ti GPU, with the PyTorch framework.

## IV. EXPERIMENTAL RESULTS

*A. Datasets Description*

We evaluated our method with the Ultra-Low Dose Imaging Challenge [23] dataset. We used the datasets that were released for the first-round challenge which consists of 398 studies where 117 studies were acquired from the Siemens Biograph Vision Quadra scanner and 281 studies were acquired from the United Imaging uEXPLORER scanner. All data were acquired in list mode allowing for the rebinding of data to simulate different acquisition times. Each simulated low statistics corresponding to *lPET* with a certain dose reduction factor (DRF) was reconstructed from the counts of a time window resampled at the middle of the acquisition with reduced time. *lPET* images were provided with DRF at 4, 10, 20, 50, and 100, as well as full-dose images. All these *lPET* were produced by subsampling a portion of the full scan, such that they are aligned with the full-dose PET. The original Siemens PET scan size is 440 × 440 × 644 with a voxel spacing of 1.65 × 1.65 × 1.65 $mm^3$, and the final acquired uEXPLORER PET image has a size of 360 × 360 × 673 with a voxel spacing of 1.667 × 1.667 × 2.886 $mm^3$.

TABLE I
QUANTITATIVE RESULTS OF EACH DRF TRAINED BY DIFFERENT COMBINATIONS OF DRF DATASETS FROM SIEMENS

| Test DATASET | \multicolumn{5}{c|}{TRAINING DATASET} | PSNR↑ (dB) | SSIM↑ | NRMSE↓ (%) |
| --- | --- | --- | --- | --- | --- | --- | --- | --- |
| | 4 | 10 | 20 | 50 | 100 | | | |
| DRF 4 | √ | | | | | 60.139 | 0.997 | 0.125 |
| | √ | √ | √ | | | **60.344** | **0.998** | **0.123** |
| | | √ | √ | √ | | 60.014 | 0.998 | 0.144 |
| | | √ | √ | √ | √ | 59.556 | 0.997 | 0.191 |
| | √ | √ | √ | √ | √ | 59.681 | 0.996 | 0.184 |
| DRF 10 | | √ | | | | 57.267 | 0.994 | 0.173 |
| | √ | √ | √ | | | 58.298 | 0.995 | **0.159** |
| | | √ | √ | √ | | **58.413** | **0.998** | 0.182 |
| | | √ | √ | √ | √ | 57.549 | 0.994 | 0.186 |
| | √ | √ | √ | √ | √ | 57.741 | 0.992 | 0.176 |
| DRF 20 | | | √ | | | 55.754 | 0.991 | 0.223 |
| | √ | √ | √ | | | 56.135 | 0.996 | 0.194 |
| | | √ | √ | √ | | **56.752** | **0.998** | **0.188** |
| | | √ | √ | √ | √ | 56.017 | 0.998 | 0.191 |
| | √ | √ | √ | √ | √ | 55.879 | 0.997 | 0.197 |
| DRF 50 | | | | √ | | 52.854 | 0.988 | 0.324 |
| | √ | √ | √ | | | 53.972 | 0.993 | 0.274 |
| | | √ | √ | √ | | **54.795** | 0.995 | **0.235** |
| | | √ | √ | √ | √ | 54.574 | **0.996** | 0.246 |
| | √ | √ | √ | √ | √ | 54.487 | **0.996** | 0.262 |
| DRF 100 | | | | | √ | 50.758 | 0.980 | 0.482 |
| | √ | √ | √ | | | 51.576 | 0.989 | 0.397 |
| | | √ | √ | √ | | 52.183 | 0.993 | 0.307 |
| | | √ | √ | √ | √ | **53.079** | 0.995 | **0.288** |
| | √ | √ | √ | √ | √ | 52.295 | **0.996** | 0.314 |

*B. Evaluation Metrics*

To assess PET synthesis performance, we adopted three evaluation metrics: Normalized root mean squared error (NRMSE), peak signal-to-noise ratio (PSNR), and structural similarity index measurement (SSIM). The higher the DRF of the *lPET*, the more challenging it is for the models to recover it to the quality of the *sPET*. Along with the evaluations on individual datasets with DRFs, an overall evaluation score was also measured where different weights were applied to each low-dose PET at different DRFs, according to:

$$Score_{avg} = w_1 * score_{DRF100} + w_2 * score_{DRF50} \\ + w_3 * score_{DRF20} + w_4 * score_{DRF10} \\ + w_5 * score_{DRF4} \quad (10)$$

where the *score* can be either of the evaluation metrics. $w_1$, $w_2$, $w_3$, $w_4$ and $w_5$ represent 35%, 25%, 20%, 15% and 5% respectively.

*C. Influence of Generalized Model*

The challenge dataset contains five individual *lPET* sub-datasets with certain dose reduction factors (DRF=4, 10, 20, 50, and 100). We employed two ways of training. Firstly, for the individual model, the network was trained with a paired set of images at the standard dose and a given DRF. The trained individual models were later tested only on corresponding DRF test datasets. Secondly, the generalized model was trained by mixing the image pairs of different DRF. Xue *et al.* [42] claimed that combining training images from multiple levels of DRF can be viewed as a data augmentation technique, which has been proved to be useful in other applications and been proven that the generalized model outperformed the individual models on cross-scanner or cross-tracer applications. However, only one combination of different DRF datasets from 4 to 20 DRF was explored in their study.

To investigate the optimal generalization capability, we combined different DRF images to train the generalized models for each DRF PET. The combination strategy is based on the proximity principle since the data distributions from the neighboring DRF are the closest. Specifically, generalized models are trained on four sub-datasets: (a) DRF 4 to DRF 20, (b) DRF 10 to DRF 50, (c) DRF 10 to DRF 100, and (d) DRF 4 to DRF 100.

The quantitative results of four generalized models and an individual model are illustrated in **Table I**. As expected, the generalized model outperformed the individual model on all DRFs. We note that the optimal dataset combination varied for different DRFs. Specifically, for *lPET* with DRF 100, the training dataset containing *lPET* with DRF 10 to 100 achieved the overall best performance with 53.079 dB PSNR and 0.288% NRMSE. However, the generalized model trained on DRF 10 to 50 resulted in the best PSNR and NRMSE with DRF 20 as the test, with 56.752 dB and 0.188% respectively, and achieves the optimal test results on DRF 50 with 54.765 dB of PSNR and 0.235% of NRMSE. Regarding the test performances of DRF 10, the generalized model trained on *lPET* data from DRF 10 to DRF 50 showed a slightly narrower margin in terms of PSBR and SSIM compared to the model trained on DRF 4 to DRF 20, with an improvement of 0.115 dB and 0.003 in PSNR and SSIM, respectively. However, the NRMSE was found to be 0.023% higher than the former model. Overall, training datasets consisting of DRF 4 to DRF 20 show the overall best results on DRF 10 testing and so does it on DRF 4 with 60.344 dB PSNR and 0.123% NRMSE.

*D. Comparison Results*

*a) Quantitative Results on Siemens Dataset*

We extensively compared our proposed SS-AEGAN with the state-of-the-art medical image synthesis methods including the top two methods on the Ultra-Low Dose challenge [23] Leader board: SF-UNet and IBRB. Our preliminary work based on SS-AEGAN was ranked third place on the challenge leaderboard. Another five advanced benchmarks were also included in our comparison. **Table II** shows the quantitative synthesis results of all synthetic models using the Siemens dataset. Our SS-AEGAN outperformed the other methods from DRF 4 to DRF 100 in all three-evaluation metrics. Only the AR-GAN, IBRN and SF-UNet improved the quality from the baseline *lPET* on all DRF datasets. SS-AEGAN surpassed AR-GAN, IBRB, and SF-UNet by 1.522dB, 1.062dB, and 0.896dB in average PNSR score. Distinct NRMSE dropping can be observed on DRF 100 from 0.857% to 0.288% with an optimal average NRMSE score of 0.170%. SS-AEGAN demonstrates moderate superiority on SSIM, which increased the value from 0.979 to 0.998 on the DRF 20 dataset with an overall best SSIM score of 0.9973.

Visual comparison results from Siemens test dataset with DRF 100 are shown in **Fig.4** (upper part) where the first row presents the transverse view of brain, the second row presents the coronal view of body, and the third row displays the transverse view of liver region. We observed that the proposed SS-AEGAN produced optimal qualitative results in visual comparison. From the second row and the third row of **Fig.4** (upper part), we can see that the synthetic *sPET* of the proposed SS-AEGAN exhibits a high level of similarity to the real *sPET*, in terms of heterogeneity of liver, image contrast, and overall structure. From the region indicated by the red circle in the first row, SS-AEGAN can recover more detailed information than other methods.

TABLE II
QUANTITATIVE COMPARISON RESULTS OF SEVEN METHODS FOR DRF 4 TO DRF 100 ON SIEMENS DATASET

| Methods | PSNR↑(dB) | | | | | | NRMSE↓(%) | | | | | | SSIM↑ | | | | | |
|---|---|---|---|---|---|---|---|---|---|---|---|---|---|---|---|---|---|---|
| | DRF 4 | DRF 10 | DRF 20 | DRF 50 | DRF 100 | AVG | DRF 4 | DRF 10 | DRF 20 | DRF 50 | DRF 100 | AVG | DRF 4 | DRF 10 | DRF 20 | DRF 50 | DRF 100 | AVG |
| Low-Dose PET | 58.718 | 54.991 | 52.156 | 48.069 | 44.484 | - | 0.154 | 0.244 | 0.343 | 0.558 | 0.857 | - | 0.999 | 0.993 | 0.979 | 0.959 | 0.948 | - |
| 3D-UNet | 52.348 | 51.077 | 50.720 | 49.866 | 48.458 | 51.138 | 0.317 | 0.348 | 0.364 | 0.402 | 0.482 | 0.355 | 0.997 | 0.996 | 0.996 | 0.995 | 0.993 | 0.9961 |
| 3D-GAN | 54.941 | 53.728 | 52.302 | 51.128 | 49.273 | 53.255 | 0.214 | 0.307 | 0.331 | 0.352 | 0.425 | 0.292 | 0.998 | 0.997 | 0.996 | 0.995 | 0.993 | 0.9967 |
| 3D-CyclGAN | 55.521 | 54.018 | 53.129 | 51.815 | 49.427 | 53.806 | 0.193 | 0.284 | 0.308 | 0.339 | 0.418 | 0.272 | 0.998 | 0.997 | 0.996 | 0.995 | 0.993 | 0.9967 |
| Stack-GAN | 57.975 | 55.937 | 54.486 | 52.304 | 49.825 | 55.510 | 0.157 | 0.206 | 0.244 | 0.312 | 0.390 | 0.222 | 0.998 | 0.997 | 0.997 | 0.995 | 0.993 | 0.9969 |
| AR-GAN | 59.223 | 56.879 | 55.085 | 52.599 | 50.837 | 56.397 | 0.141 | 0.187 | 0.231 | 0.312 | 0.382 | 0.208 | 0.998 | 0.997 | 0.997 | 0.996 | 0.994 | 0.9971 |
| IBRB | 58.496 | 57.236 | 56.217 | 54.583 | 52.870 | 56.857 | 0.144 | 0.172 | 0.195 | 0.240 | 0.294 | 0.183 | 0.997 | 0.997 | 0.997 | 0.996 | 0.995 | 0.9968 |
| SF-UNet | 58.819 | 57.464 | 56.341 | 54.488 | 52.578 | 57.023 | 0.140 | 0.168 | 0.194 | 0.244 | 0.306 | 0.182 | 0.997 | 0.997 | 0.997 | 0.996 | 0.995 | 0.9968 |
| **SS-AEGAN** | **60.344** | **58.298** | **56.752** | **54.795** | **53.079** | **57.919** | **0.123** | **0.159** | **0.188** | **0.235** | **0.288** | **0.170** | 0.998 | 0.997 | **0.998** | 0.996 | 0.995 | **0.9973** |

*b) Quantitative Results on uEXPLORER Dataset*

We also note that SS-AEGAN outperformed other methods on PSNR and NRMSE measurements. Specifically, SS-AEGAN shows noticeable improvements when compared with commonly used model, 3D-UNet by raising the overall PSNR score from 50.182dB to 57.367dB and dropping the average NRMSE from 0.744% to 0.330%. Although SF-UNet slightly outperforms SS-AEGAN with an overall SSIM performance by 0.004, the proposed method shows relatively higher advantages on other two indicators by increasing PNSR by 0.509dB and decreasing NRMSE from 0.189% to 0.175%.

**Table II** and **Table III** show that the *lPET* images with DRF 50 and DRF 100 from uEXPLORER exhibit lower PSNR values (47.206dB and 42.490dB) and higher NRMSE values (0.597 and 1.402) compared to the corresponding *lPET* images from the Siemens datasets (48.069dB and 44.484dB for PSNR; 0.558 and 0.857 for NRMSE. Despite the more challenging situation posed by the uEXPLORER dataset, SS-AEGAN recovered *lPET* with DRF 50 and DRF 100 from uEXPLORER dataset to 54.024dB and 52.013dB with an overall PSNR score of 57.367dB which is comparable to the quality of the synthesized *sPET* derived from the Siemens dataset. Although the second-best method, SF-UNet achieved relatively consistent performance on uEXPLORER and Siemens datasets with DRF 50 (54.488dB and 53.377dB) and DRF 100 (52.578dB and 51.032dB), the cross-scanner performance gap is larger compared to the proposed SS-AEGAN.

**Fig.4** (lower part) shows the example of visual comparison results acquired from the uEXPLORER test set. We observe that our method outperformed all other comparison methods regarding enhancing image contrast, shown as the texture of liver in the third row of **Fig.4** (lower part) and preserving structural information e.g., spine and organs shown in the second row of **Fig.4** (lower part). Based on the regions highlighted by the red circle in **Fig.4** (lower part), it becomes apparent that the synthetic images produced by SS-AEGAN closely resembled the real *sPET*, exhibiting a higher degree of preservation of detailed information when compared to other synthesized results.

To verify the significance of the observed improvements of proposed SS-AEGAN, we conducted paired t-test between rival results and our results on both datasets. As reported in the **Table S1** and **S2**, most of the p-values are less than 0.05, indicating statistical significance in the performance improvement achieved by our method. Details are illustrated in the supplementary materials.

*E. Cross-scanner and Cross-tracer Generalizability*

To validate the cross-scanner generalizability of the proposed SS-AEGAN, a series of experiments were conducted in which the model was trained with data from one scanner and then applied to data from another scanner, shown in **Table IV** where D1 denotes uEXPLORER and D2 denotes Siemens.

The PSNR values are generally higher when the training and testing datasets are the same (D1-D1 and D2-D2), indicating that the model performs better when trained and tested on data from the same scanner. However, even when the training and testing datasets are different (D1-D2 and D2-D1), the PSNR decreases less than 1dB for DRF4 to DRF 10 and less than 2dB for DRF50 to100, suggesting that the model is capable of generalizing to new scanners.

The uEXPLORER dataset has three different tracers, FDG (259 cases), DOTA (7 cases), and 68-Ga (15 cases). DOTA and 68-GA only are used in the test stage, the diverse-tracer dataset - uEXPLORER, and the single-tracer dataset – Siemens achieved comparable performance on three metrics as illustrated in **Table II** and **Table III**. The cross-tracer visualization examples are shown in **Fig.5**, all three tracer types consistently show that the synthesized *sPET* images of DRF4 to DRF20 closely resemble the true *sPET* images, especially in high-intake regions. Despite the poor image quality of input *lPET*, which inevitably leads to some missing information in the synthetic results of DRF50 and DRF100, SS-AEGAN is still able to recover most of the structures and high-intake regions in all three types of tracer PET.

TABLE III
QUANTITATIVE COMPARISON RESULTS OF SEVEN METHODS FOR DRF 4 TO DRF 100 ON uEXPLORER DATASET

| Methods | PSNR↑(dB) | | | | | | NRMSE↓(%) | | | | | | SSIM↑ | | | | | |
|---|---|---|---|---|---|---|---|---|---|---|---|---|---|---|---|---|---|---|
| | DRF 4 | DRF 10 | DRF 20 | DRF 50 | DRF 100 | AVG | DRF 4 | DRF 10 | DRF 20 | DRF 50 | DRF 100 | AVG | DRF 4 | DRF 10 | DRF 20 | DRF 50 | DRF 100 | AVG |
| Low-Dose PET | 58.718 | 54.421 | 51.698 | 47.206 | 42.490 | - | 0.154 | 0.249 | 0.347 | 0.597 | 1.402 | - | 0.999 | 0.997 | 0.995 | 0.983 | 0.970 | - |
| 3D-UNet | 52.871 | 49.481 | 49.231 | 48.149 | 44.770 | 50.182 | 0.317 | 0.424 | 0.436 | 0.501 | 0.744 | 0.417 | 0.995 | 0.993 | 0.994 | 0.993 | 0.992 | 0.9939 |
| 3D-GAN | 54.487 | 52.315 | 51.827 | 48.745 | 45.904 | 52.122 | 0.284 | 0.319 | 0.343 | 0.426 | 0.625 | 0.343 | 0.996 | 0.994 | 0.995 | 0.994 | 0.993 | 0.9949 |
| 3D-CyclGAN | 55.084 | 53.781 | 52.418 | 49.527 | 47.282 | 53.001 | 0.240 | 0.286 | 0.307 | 0.398 | 0.495 | 0.301 | 0.997 | 0.994 | 0.995 | 0.995 | 0.993 | 0.9954 |
| Stack-GAN | 57.017 | 55.677 | 53.709 | 50.832 | 49.074 | 54.696 | 0.162 | 0.210 | 0.264 | 0.373 | 0.421 | 0.239 | 0.998 | 0.998 | 0.997 | 0.994 | 0.993 | 0.9970 |
| AR-GAN | 58.794 | 56.739 | 55.379 | 53.151 | 50.674 | 56.345 | 0.138 | 0.178 | 0.211 | 0.278 | 0.377 | 0.196 | 0.998 | 0.998 | 0.998 | 0.996 | 0.994 | 0.9975 |
| IBRB | 58.630 | 56.846 | 55.635 | 53.606 | 51.437 | 56.472 | 0.140 | 0.175 | 0.203 | 0.261 | 0.341 | 0.190 | 0.998 | 0.998 | 0.998 | 0.996 | 0.994 | 0.9975 |
| SF-UNet | 59.714 | 57.139 | 55.607 | 53.337 | 51.032 | 56.858 | 0.130 | 0.173 | 0.207 | 0.271 | 0.358 | 0.189 | 0.999 | 0.999 | 0.998 | **0.997** | 0.994 | **0.9983** |
| **SS-AEGAN** | **60.160** | **57.457** | **56.214** | **54.024** | **52.013** | **57.367** | **0.123** | **0.167** | **0.182** | **0.251** | **0.330** | **0.175** | 0.999 | 0.999 | 0.998 | 0.995 | 0.993 | 0.9979 |

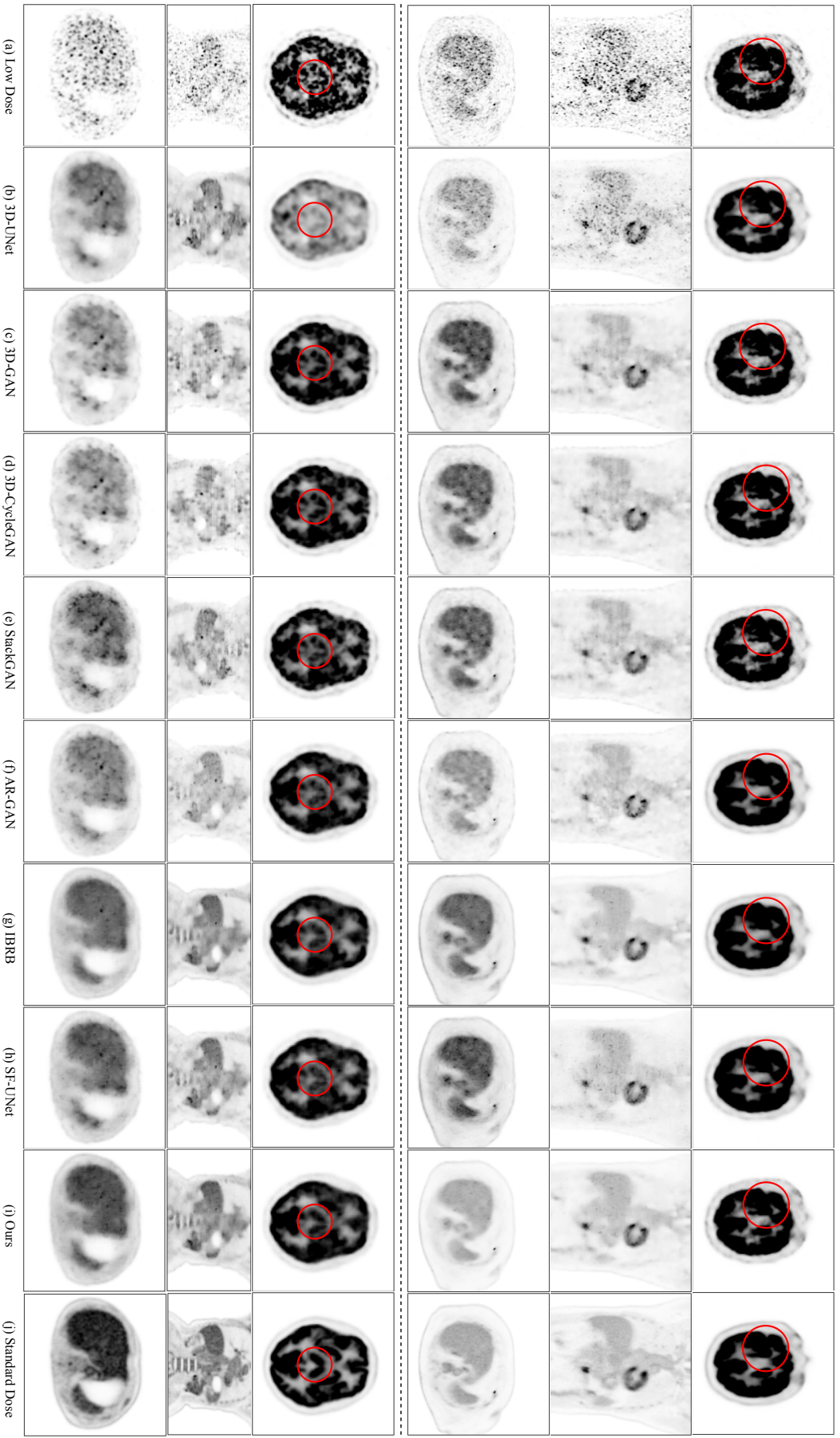

**Fig. 4.** Visual comparison results of various methods on test set with DRF100 from uEXPLORER (Row 1-3) and Siemens (Row 4-6). Rows are different selected views. (1) Transverse view of brain region. (2) Coronal view. (3) Transverse view of liver region.

TABLE IV
CROSS-SCANNER GENERALIZATION STUDY.

| Train | Test | PSNR (dB)↑ | | | | |
|---|---|---|---|---|---|---|
| | | DRF4 | DRF10 | DRF20 | DRF50 | DRF100 |
| D1 | D1 | 60.344 | 58.298 | 56.752 | 54.795 | 53.079 |
| D2 | D1 | 59.687 | 57.463 | 55.694 | 53.840 | 51.957 |
| D2 | D2 | 60.160 | 57.457 | 56.214 | 54.024 | 52.013 |
| D1 | D2 | 59.246 | 56.594 | 55.459 | 52.793 | 50.991 |

*D1 denotes uEXPLORER, D2 denotes Siemens.

TABLE V
ABLATION STUDY OF 3D-AEGAN MODULE ON TEST RESULTS
OF DRF 100 FROM D1-UEXPLORER AND D2-SIEMENS.

| | Pix | AE | Dis | PSNR↑ | SSIM↑ | NRMSE↓ |
|---|---|---|---|---|---|---|
| | Low-Dose | | | 42.490 | 0.970 | 1.402 |
| | √ | | | 46.019 | 0.990 | 0.475 |
| D1 | √ | √ | | 49.175 | 0.991 | 0.361 |
| | √ | | √ | 48.314 | 0.991 | 0.441 |
| | √ | √ | √ | **51.288** | **0.992** | **0.352** |
| | Low-Dose | | | 44.484 | 0.948 | 0.857 |
| | √ | | | 48.878 | 0.992 | 0.479 |
| D2 | √ | √ | | 50.923 | 0.993 | 0.313 |
| | √ | | √ | 49.617 | 0.994 | 0.443 |
| | √ | √ | √ | **51.316** | **0.994** | **0.301** |

*Pix, AE, and Dis denote Pixel-Net, AE-Net, and Discriminator respectively.

*F. Ablation Study*

To evaluate the effectiveness of individual components of our SS-AEGAN, we conducted multiple ablation studies on self-supervised pre-training strategy and 3D-AEGAN module with DRF 100.

*a) Synthesis Network-3DAEGAN:*

To investigate how each component of 3D-AEGAN improves the synthetic image quality, we decoupled the 3D-AEGAN into (a) baseline, synthesis results of Pixel-Net; (b) baseline with adaptive residual estimation network (Pix + AE); (c) baseline with discriminator (Pix + Dis); and (d) baseline with AE-Net and discriminator (Pix + AE + Dis).

As shown in the first row and second row in **Table V**, the baseline Pixel-Net improved *lPET* of uEXPLORER from 42.490 dB to 46.019 dB on PSNR; increased SSIM from 0.970 to 0.990 and dropped NRMSE from 1.402% to 0.475%. For *lPET* of Siemens, Pixel-Net increased image quality by 4.394 dB on PSNR, 0.044 on SSIM, and 0.378% NRMSE correction.

To verify the residual mapping ability of the proposed AE-Net, we compared the results between baseline with/without AE-Net. The quantitative results are shown in the second and third rows of **Table V**. The AE-Net improved the baseline Pixel-Net by 3.156 dB, 0.001, and 0.114% in PSNR, SSIM, and NRMSE respectively on DRF 100 from uEXPLORER. Similarly, AE-Net also shows its advantages on Siemens datasets by improving the synthetic results of Pixel-Net by 2.045 dB, 0.001, and 0.166% in PSNR, SSIM, and NRMSE. It indicated that the proposed AE-Net was effective for high-quality PET synthesis, especially in spatial information recovery.

To further assess the performance of AE-Net, we compared the results between the baseline with AE-Net and the baseline with discriminator, shown as the third row and fourth row of two sub-tables in **Table V**. The baseline with the AE-Net outperformed the baseline with the discriminator by 0.861dB and 1.306 dB in PSNR on uEXPLORER and Siemens respectively. Similarly, AE-Net with baseline noticeably surpassed baseline with discriminator on the pixel-wise correction and decreased NRMSE from 0.441% to 0.361% on uEXPLORER and from 0.443% to 0.313% on Siemens. It indicates that the proposed AE-Net is a powerful tool for high-quality image synthesis, especially for very-low-dose PET images e.g., DRF 100.

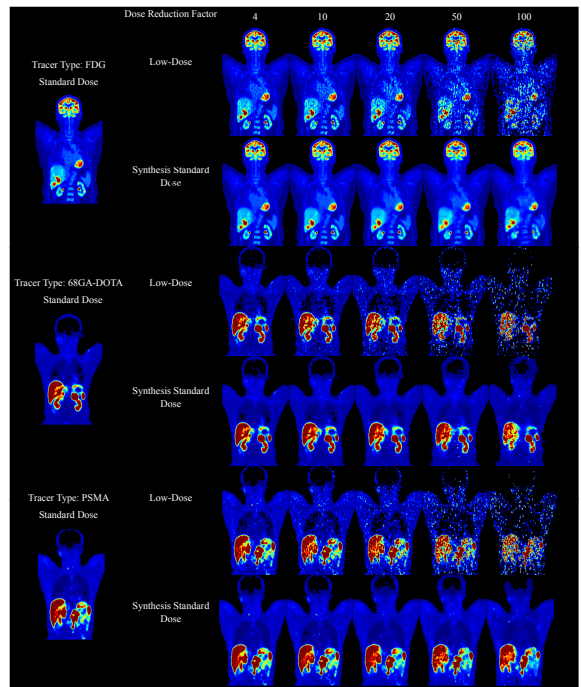

**Fig. 5.** Examples of FDG PET, 68-GA DOTA PET and PSMA-PET from uEXPLORER test set with low-dose PET and the corresponding synthesized standard PET.

By adding a discriminator into the baseline + AE-Net, PSNR improved to 51.288dB from 49.175dB and NRMSE dropped to 0.352% on uEXPLORER; PSNR increased to 52.316dB and NRMSE decreased to 0.301% on Siemens. Making a synthesis model into a GAN-similar structure showed the advantages of structural level recovery by increasing SSIM to 0.992 with the uEXPLORER scanner and to 0.994 with the Siemens scanner.

TABLE VI
ABLATION STUDY OF THE EFFECTIVENESS OF EACH
UPSTREAM TASK IN SELF-SUPERVISED PRE-TRAINING
STRATEGY FOR DRF 100.

| Up-stream Tasks | Average Score | | |
|---|---|---|---|
| | PSNR↑ | SSIM↑ | NRMSE↓ |
| Train from Scratch | 51.288 | 0.992 | 0.352 |
| $L_{CPC}$ | 51.437 | 0.992 | 0.347 |
| $L_{rotation}$ | 51.326 | 0.992 | 0.350 |
| $L_{resotoration}$ | 51.752 | 0.993 | 0.342 |
| $L_{classification}$ | 51.557 | 0.993 | 0.343 |
| $L_{CPC} + L_{rot} + L_{res} + L_{class}$ | **52.013** | **0.993** | **0.330** |

*b) Efficacy of Self-Supervised Pre-training:*

**Fig.8** shows the comparison results of using self-supervised pre-training (SSP) when compared to training from scratch. A notable improvement at PSNR can be observed for DRF 100, raising from 51.288dB to 52.013dB for uEXPLORER and from 52.316dB to 53.079dB for Siemens. The second largest gap shows in DRF 50, the SSP strategy improved both test results by 0.716dB and 0.678dB for uEXPLORER and Siemens respectively. The test results of DRF 20 and DRF 10 demonstrate substantial improvements, as the SSP strategy increased PSNR by 0.653 dB and 0.443 dB for uEXPLORER and boosted PSNR by 0.541 dB and 0.452 dB for Siemens. Even with DRF 4, minor performance optimization by SSP can still be observed on both scanners.

TABLE VII
QUANTITATIVE COMPARISON RESULTS OF COMBINATION OF 3D-GAN WITH PROPOSED MODULE ON uEXPLORER DATASET WITH DRF 100.

| Method | AE-Net | SSP | PSNR↑ | NRMSE↓ | SSIM↑ |
|---|---|---|---|---|---|
| 3D-GAN | | | 45.904 | 0.625 | 0.993 |
| 3D-GAN | √ | | 47.383 | 0.537 | 0.992 |
| 3D-GAN | | √ | 46.218 | 0.612 | 0.993 |
| 3D-GAN | √ | √ | 48.014 | 0.531 | 0.993 |

*c) Efficacy of Self-Supervised Up-stream Tasks:*

To investigate the impact of individual upstream tasks in the self-supervised pre-training strategy, we conducted multiple experiments as shown in **Table VI**.

The self-restoration task-guided self-pretraining achieved the best performance by improving 0.464 dB in PSNR and 0.01% in NRMSE. When employing all the up-streams tasks in the SSP, it obtains the optimal test results with PSNR improvement of 0.725dB.

### G. Clinical Assessment on Liver ROIs

To evaluate the robustness of the synthesized results, we measured the homogeneity of a region of interest (ROI) in the liver structure. Such ROI measurement can be used to quantify image quality as sections of the liver is expected to be homogeneous [53]. We manually annotated spherical regions of interest (ROIs) with a diameter of 20±1 mm within lesion-free and homogeneous sections of the right liver lobe [52]. Our annotation process avoided sections of the liver that includes prominent blood vessels and partial volume effect. Fig.S1 illustrates a case of this annotation process. We conducted an analysis involving both the rival methods and our method where we measured SUVmax and SUVmean. We calculated the accuracy of SUVmax and SUVmean within ROIs in reference to the standard dose PET, quantifying this accuracy in terms of percentage error. This evaluation was conducted using the uEXPLORER dataset with DRF 100 among all 28 patient studies. This data was selected as the uEXPLORER dataset has more cases compared to 11 cases of Siemens dataset and DRF 100 represents the most challenging cases.

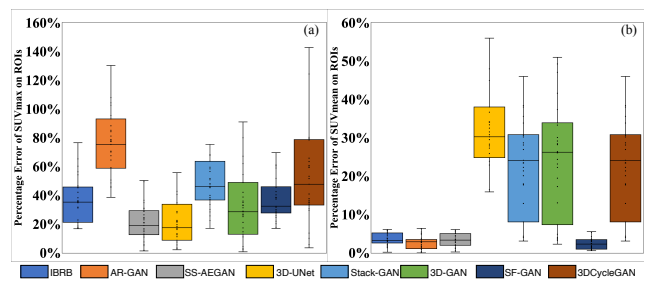

**Fig.6** Percentage error of SUVmax (a) and SUVmean (b) of uEXPLORER test dataset with DRF100.

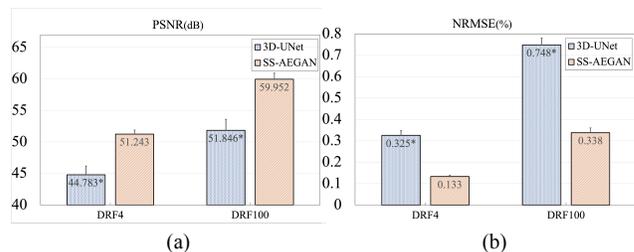

**Fig.7** Quantitative comparison of 5-fold cross validation results on PSNR (a) and NRMSE (b), a significant difference compared with the proposed SS-AEGAM is indicated by * (p < 0.005).

The percentage errors for SUVmax and SUVmean are visually presented in Fig.6. Notably, our SS-AEGAN achieved the lowest percentage error in terms of SUVmax. Additionally, our method ranks as the second-best in SUVmean. These results underscore the robustness of our method in comparison to rival methods.

### H. Model Variability Assessment

To validate the stability of our model performance, we implemented a 5-fold cross-validation during training and validation. Individual tests were conducted on a separate test dataset to ensure a fair comparison. We used DRF 4 and DRF 100 from the uEXPLORER dataset to represent the lowest dose and highest dose reduction factors, respectively. The baseline model 3D-UNet was used as the comparison method. Quantitative results are presented in Fig.7 in terms of PSNR and NRMSE, demonstrating that our SS-AEGAN exhibits lower standard deviations, and the observed superior performance is statistically significant, as indicated by the low p-values associated with the results.

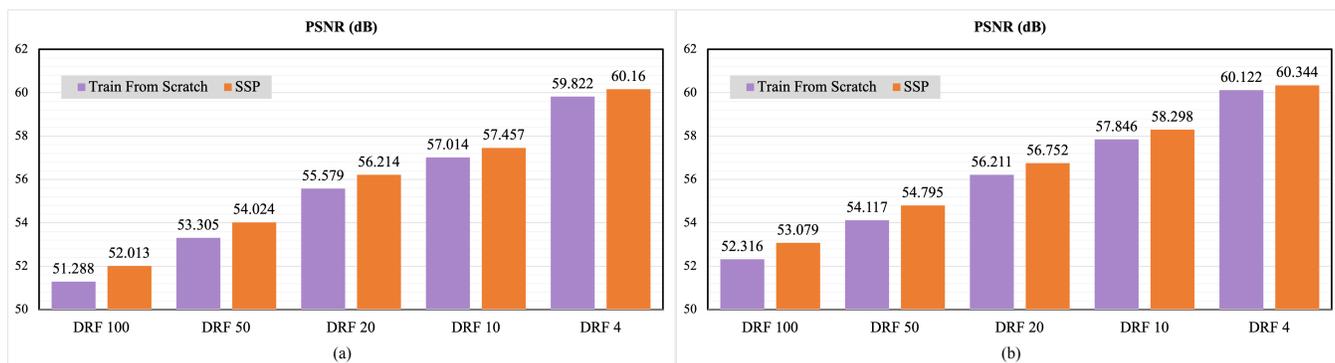

**Fig. 8.** Ablation study between using self-supervised pre-training (SSP) strategy and training from scratch strategy tested on (a) uEXPLORER dataset with DRF 100 to DRF 4 and (b) Siemens dataset with DRF 100 to DRF 4.

V. DISCUSSION

Our main findings are that: (1) our SS-AEGAN consistently outperformed the state-of-the-art methods, in particular, for DRF 100, where the image characteristics suffered from loss of structure information and usually presented low signal-to-noise ratio; (2) we identified that our method has promising generalizability through cross-scanner, cross-tracer and cross model analysis; (3) self-supervised pre-training increases discrimination power of the derived features; and (4) input-involved residual estimation can effectively narrow down the difference between synthesis *sPET* and *lPET*.

**Table II** and **Table III** show that our method achieved the best results across the two scanners. The improvement of 3D-GAN over 3D-UNet is likely attributed to the use of adversarial learning; adversarial learning has a discriminator that allows it to distinguish between the synthesized and real images, enabling it to generate higher quality images when compared with the 3D-UNet structure. The further improvement of 3D-CyclGAN is attributed to the cycle consistency loss which encourages the generator to produce output images that can be mapped back to the original input domain. StackGAN achieved better performance than CycleGAN, this was attributed to its multi-stage architecture, which generates images in multiple resolutions and incorporates a conditioning auxiliary variable at each stage. This approach allows StackGAN to generate high-quality images with greater anatomical detail. However, StackGAN still suffers from feature representation limitations in PET image generation due to the reuse of the same network structure in each stage of the generator. AR-GAN effectively resolves this issue by using the residual estimation module as the second stage component to refine the synthetic output from the first stage. Both SF-UNet and IBRB achieved competitive performance when compared to our proposed method. However, both trained a single-stage model from scratch which may limit feature representation in generating fine-grained details and in recovering complex textures e.g., tumors with inhomogeneous textures. In contrast, our method SS-AEGAN can minimize these limitations by adopting self-supervised pre-training with a second-stage residual estimation module - AE-Net.

We also investigated the effect of trainable parameters against all comparison methods. Fig.S2 and Fig.S3 present the comparisons. The SS-AEGAN is third least in the number of parameters used yet outperformed all comparison methods, in both PSNR and NRMSE, even to methods that used three times more parameters.

For the cross-scanner and cross-tracer evaluation, the results in **Table IV** indicate that the SS-AEGAN has promising generalizability across different scanners. We suggest that the model's generalizability is mainly attributed to the adopted self-supervised pre-training strategy (SSP). First, using the self-supervised method that involves inpainting (such as self-restoration) is expected to improve a model's ability to handle noisy or incomplete data during inference. Contrastive coding can further aid a model to learn useful and robust features that are generalizable across different datasets or tasks.

Compared with models trained from scratch, the SSP model shows overall better synthesis performance on all DRFs. The proposed four upstream self-supervised pre-training tasks boost the feature representation by learning multidimensional information, and the individual influence of each task can be observed in **Table VI**. The classification task involves training the model to predict the dose reduction level of an input image. This task can encourage the model to learn more abstract and higher-level features that are invariant to changes in appearance or context. In the context of PET image synthesis, we identified that this process could help the trained model to better capture the underlying characteristics of PET images that are relevant to the synthesis task.

Generalizability of proposed method also shows in the integrability of the proposed AE-Net and SSP components. AE-Net and SSP are claimed to be easily incorporated with any synthesis generator and likely be able to boost the synthesis results. We conducted an additional ablation study by combining the proposed AE-Net and SSP with commonly used generator, 3D-GAN, with the results shown in Table VII. The outcomes illustrate the effectiveness of our individual component on improving synthesis outcomes.

The proposed AE-GAN assisted in high-quality PET synthesis as shown in **Table V**. Compared to the current residual estimation method AR-GAN [33], the advantages of AE-GAN were as follows. The previous residual estimation method used synthetic PET as input to predict the difference map between synthetic PET and *sPET*. However, due to domain unalignment between synthetic PET and *sPET*, synthetic PET as input may not be capable of providing all the information needed to estimate the residual. In contrast, we replaced the single-dimensional input, synthetic PET, with two-dimensional input, the difference map between *lPET* and synthetic PET, to mimic the residual between low-dose PET and standard-dose PET. The input-involved residual estimation method compensates for incomplete information in the synthetic image and bridges the domain gap between *lPET* and *sPET* in a more direct manner. Further, when compared to the 2D method in [33], AE-GAN using a 3D adaptive residual estimation method was able to capture spatial dependencies and structural information along the three views.

In this study, we mainly focused on introducing the difference in residual map for reconstructing the standard dose PET images. Therefore, a standard additive residual scheme was used. In our future work, we will compare different residual scheme such as the multiplicative residual scheme proposed by Guo *et al.* [34] further evaluate the performance of the proposed method.

Besides, by leveraging recent advancements in AI-driven reconstruction methods [9-11] that directly reconstruct high-quality PET images from low-dose sinograms, we aim to explore the incorporation of the end-to-end structure of our model, which inherently possesses the capability to take sinograms as input, into the domain of high-quality PET reconstruction from low-dose sinograms.

ACKNOWLEDGMENT

All authors declare that they have no known conflicts of interest in terms of competing financial interests or personal relationships that could have an influence or are relevant to the work reported in this paper.

VI. CONCLUSION

In this paper, we designed a self-supervised pre-trained adaptive residual estimation-based generative adversarial network (SS-AEGAN) for high-quality standard-dose PET synthesis from low-dose PET. To enhance the model generalizability and feature representation, a self-supervised pre-training is introduced with four up-stream tasks to assist high-quality *sPET* synthesis. Moreover, a novel *lPET*-involved residual estimation module is proposed to further narrow the distribution misalignment between *sPET* and *lPET*. Experimental results with a large public benchmark dataset demonstrated that our method surpassed the current state-of-the-art methods. As part of our future work, we plan to incorporate prior knowledge e.g., anatomy derived from CT, and MRI to further improve the quality and sensitivity of the synthesized PET images.